\documentclass{sf2a-conf2011}
\usepackage{graphicx}
\usepackage{hyperref}
\usepackage[]{natbib}  
\usepackage[cyr]{aeguill}
\usepackage{epstopdf}

\def\BibTeX{{\rm B\kern-.05em{\sc i\kern-.025em b}\kern-.08em
    T\kern-.1667em\lower.7ex\hbox{E}\kern-.125emX}}
\bibpunct{(}{)}{;}{a}{}{,}  

\begin{document}

\TitreGlobal{SF2A 2011}


\title{Surface convection: from the Sun to red giant stars}

\runningtitle{Solar and RGB surface convection}

\author{L. Piau}\address{LATMOS, 11 Boulevard d'Alembert, 78280
 Guyancourt, France}

\author{P. Kervella}\address{LESIA, Observatoire de Paris, CNRS\,UMR\,8109, UPMC, Universit\'e Paris Diderot, 5 place
Jules Janssen, 92195 Meudon, France}

\author{S. Dib}\address{Astrophysics Group, Imperial College of Science, Technology and Medicine, London SW7 2AZ, United Kingdom}

\author{P. Hauschildt}\address{Hamburger Sternwarte, Gojenbergsweg 112, 21029 Hamburg, Germany}


\setcounter{page}{237}

\index{Author1, L.}
\index{Author2, P.}
\index{Author3, S.}
\index{Author4, P.}


\maketitle


\begin{abstract}

We check how the change in surface conditions between the Sun and 
red giant branch stars changes the characteristic surface convection
length scale to be used in models. We investigate the question in the
case of the mixing length theory and of the phenomenology of full
spectrum of turbulence. For the observational part, we rely on 
independent measurements of effective temperatures and interferometric
radii of nearby red giants. 
We find that the local red giant branch cannot be explained taking
into account the solar calibrated convective length scale.

\end{abstract}



\section{Introduction}\label{sec1}

In low mass stars the effective temperature and the radius estimate 
the efficiency of surface convection. The Sun and the red giant branch
stars (hereafter RGBs) have 
very different surface conditions. A red giant surface gravity and 
energy flux are much smaller than the solar ones. The purpose of the work 
we present is to check how this changes the surface convection efficiency.
First we build calibrated solar models using the mixing length theory 
(hereafter MLT) or the phenomenology of full spectrum of turbulence by 
Canuto, Goldman \& Mazzitelli (1996) (hereafter CGM) for surface convection. 
Then relying on the same input physics, we build red giant models. The 
red giant models radii and effective temperatures are compared to an 
observational sample of 38 objects for which the radii are known directly 
through interferometry to better than 10 percent. Absolute luminosities 
and effective temperatures of these objects are also accurately known. In 
the next section we give the main ingredients of the models affecting the 
radius and effective temperature and address 
the solar calibration. In section \ref{sec3} we 
describe the RGBs giant sample. Section \ref{sec4} investigates the 
changes in characteristic convection length scale from the Sun to RGBs. 
We conclude in section \ref{sec5}.

\section{Models inputs and solar convection length scale}\label{sec2} 

The radii of solar and RGBs models are tuned by the outer
thermal gradient. It depends on four ingredients: 

i) The opacities. We rely on the OPAL opacities and, below 5600K, 
on the \cite{ferguson05} opacities. The composition is either
 assumed to be solar with X=0.7392, Z=0.0122 (\cite{Asplund05}) 
or slightly subsolar at [Fe/H]=-0.17. The metal repartition is always 
the one of \cite{Asplund05}.  

ii) Convection efficiency to transport energy. The poor efficiency of 
outer convection induces the subsurface superadiabatic gradient which 
in turn sets the entropy of the deep convection zone. A lower entropy 
level of the deep convection zone means a less dense envelope and thus 
a wider radius and a lower effective temperature. To model the inefficient 
convection we use two simplified local treatments of convection: the mixing 
length theory (\cite{Bohm58}) and the full spectrum of 
turbulence of \cite{Canuto96}. For both 
treatments the characteristic convection length scale $\rm \Lambda$ is assumed 
to be a constant fraction of the local pressure scale height: 
$\rm \Lambda= \alpha H_p$.

iii) Atmospheric effects. At the edge of the star the diffusion 
approximation does not hold for photons and lines strongly 
affect the radiative transfer: these effects are expressed by the 
temperature-optical depth  relations that are provided by atmosphere models.
We use two series of non-grey atmosphere models as outer boundary conditions. 
The first series of relations ($\rm T(\tau)^4=T_{eff}^4 f_{grid}(\tau)$) is
computed with the PHOENIX/1D atmosphere code where the convection is 
handled using the MLT. The second series of temperature-optical depth 
relations is computed with the Atlas12 atmosphere code 
(\cite{castelli05}). We modified Atlas12 to use the CGM 
prescription. Each type of atmosphere
 models are used with the corresponding phenomenology of convection in 
the deeper regions as is necessary for consistency in 
the models (\cite{montal04}).

iv) The equation of state influences the radius through
 the adiabatic exponents. We use the OPAL EoS.

We assume that $\rm L_{\odot}=3.846\,10^{33} erg.s^{-1}$ and 
$\rm R_{\odot}=6.9599\,10^{10} cm$ and begin the solar evolution
 on the zero age main sequence. The calibration in luminosity, radius,
 and metal-to-hydrogen ratio $\rm \frac{Z_{surf}}{X_{surf}}$ are achieved 
to better than $10^{-4}$ at the age of 4.6 Gyr for both MLT and CGM
 convection prescriptions. In the MLT framework we obtain 
$\rm \alpha_{mlt\odot}$=1.98, in the CGM framework we obtain 
$\rm \alpha_{cgm\odot}$=0.77.

\section{The red giant sample}\label{sec3}

We first queried the CHARM2 catalogue (\cite{richichi05}) 
to obtain all direct measurements of giant and subgiant angular diameters
 up to 2004, with effective temperatures in the range from 5000\,K to 5500\,K.
 We then searched the literature for more recent observations, and added the 
measurement of $\gamma$\,Sge, $\delta$\,Eri, $\xi$\,Hya, and the recent high 
accuracy CHARA/FLUOR measurements of $\epsilon$\,Oph and $\eta$\,Ser. The 
conversion of uniform disk angular diameters to limb-darkened values was 
done using linear limb-darkening coefficients by \cite{claret95}, 
which are based on stellar atmosphere models by \cite{kur93}.
Our sample contains 38 giant and subgiant stars with spectral types from G5 
to M0. The distances to the selected stars range from 11 to 110\,pc. Thanks 
to this proximity, we neglected the interstellar reddening for the 
computation of the bolometric luminosity. The accurate parallaxes and 
interferometric angular radii estimates allow the objects to have a relative 
uncertainty in the linear radius smaller than 10\%. The average metallicity 
is slightly subsolar $\rm [Fe/H]= - 0.17$ with no object below -0.44 and no 
object above 0.04 but one exception at 0.13. 

\section{Red giant branch calibration}\label{sec4}

\begin{figure}[Ht!]
 \centering
 \includegraphics[width=0.30\textwidth,angle=90,clip]{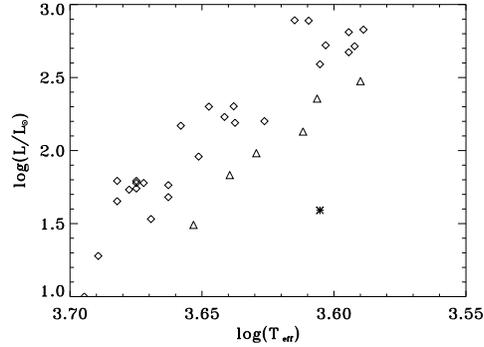} 
  \caption{HR diagram of the RGB stars sample. The triangles define 
the lower envelope of the RGB in the sense that they show the stars with the 
lowest effective temperature for a given luminosity.}
  \label{fig1}
\end{figure}

We model RGB stars up to $10^3$ solar luminosity with exactly the same 
physics as in the solar models. The microscopic diffusion is accounted for 
in any model warmer than 5000K following \cite{proffitt93}. 
This is important in order to obtain correct 
ages as diffusion speeds up the main sequence evolution. After the first 
dredge-up though ($\rm T_{eff}< 5000K$) diffusion effects 
become negligible. Provided 
the atmosphere boundary models and the opacity tables are unchanged, there 
are three main models inputs that change the position of the RGB: the mass, 
the metallicity and the surface convection characteristic length scale. The 
latter parameter is what we want to constrain. As mentioned above the more 
efficient the convection, the smaller the radius and the higher the effective 
temperature at a given luminosity. The metallicity of the sample is known and 
is therefore no hurdle. The masses of the stars however are not known: unlike 
RGB stars of a globular or a Galactic cluster, the RGB stars of the sample 
are field stars that presumably have different masses and ages. 
Yet it is possible to set an upper limit to the age of local red giants or 
equivalently a lower limit to their masses. The limit is given by the age of 
the Galactic disk and the evolutionary timescale of its low mass stars: for 
objects in the slightly subsolar metallicity range 
($\rm -0.25 <[Fe/H] < -0.14$), \cite{liu00} suggest a 
maximum age of $\rm 11.7\pm 1.9$ Gyr. In this study, we consider models that 
have reached $\rm 10^3 L_{\odot}$ on the RGB by that age as our RGB stars 
exhibits $\rm [Fe/H]=-0.17$ on average. We can set broader upper limits to 
the local Galactic disk age: it is certainly younger than the Universe 
$\rm 13.7 \pm 0.1$ Gyr (\cite{komatsu09}). In the next 
subsections we focus on the lower envelope of the RGB i.e. on these stars 
with the lower effective temperature or larger radii at a given luminosity 
(see Figure~\ref{fig1}). They are also the oldest and lowest mass stars of 
the sample.

\subsection{Mixing length theory}\label{41}

Figure~\ref{fig2} left panel shows the six lower envelope stars of the RGB and 
features different evolutionary tracks. Let's define 
$\rm \chi_1^2=\sum_{i=1}^{N}\frac{1}{N}[\frac{T_{eff}^{mod}-T_{eff}^{obs}}{\Delta T_{eff}^{obs}}]^2$, 
where $\rm T_{eff}^{obs}$, $\rm T_{eff}^{mod}$ and $\rm \Delta T_{eff}^{obs}=130K$
 are respectively the observed $\rm T_{eff}$ of an object, the $\rm T_{eff}$ of 
the model having the same luminosity as the object and the uncertainty on the 
observed $\rm T_{eff}$. N is the number of objects considered. 
Figure~\ref{fig2} left panel models in solid line, dashed line and dotted line all use 
the solar calibrated value of $\rm \alpha_{mlt}=1.98$. The solid line track 
corresponds to a 0.95$\rm M_{\odot}$ star reaching $\rm 10^3 L_{\odot}$ at 
$\rm \approx$11.5 Gyr. This model has $\rm [Fe/H]=-0.17$ and an helium fraction 
Y=0.2582. It is clearly too warm to fit the lower envelope of the RGB: 
$\rm \chi_1^2$=3.8.  The dashed and dotted lines correspond to less massive and 
helium poorer stars with 0.9$\rm M_{\odot}$ and Y=0.2482 respectively. The 
lower mass model is extreme in the sense that it reaches $\rm 10^3 L_{\odot}$ 
at $\approx$13.9 Gyr (or $\rm 10^2 L_{\odot}$ at $\approx$13.88 Gyr), which is 
older than the current age estimate of the Universe. The helium poor model is 
also extreme in the sense that its helium fraction nearly is that of Big Bang 
nucleosynthesis (\cite{coc04}) and evidently cannot be lowered any 
further. Both two last models are in slightly better agreement with the data 
than the former one. Yet they do not provide a good fit to the observations. 
This demonstrates that mass or helium fraction cannot be changed to improve 
the agreement to the observations.
Models with lower $\rm \alpha_{mlt}$ than the solar value provide much better 
fits to the data. The three dotted-dashed line is the track of the model 
with $\rm \alpha_{mlt}=1.68$, 0.95$\rm M_{\odot}$, and $\rm [Fe/H]= -0.17$. This 
model reaches $\rm 10^3 L_{\odot}$ at 11.5 Gyr and has $\chi_1^2$=0.16. The 
long-dashed line model has $\rm \alpha_{mlt}=1.68$, 1.13$\rm M_{\odot}$, and 
$\rm [Fe/H]=0$. It reaches $\rm 10^3 L_{\odot}$ at 7.5 Gyr and has 
$\chi_1^2$=0.078. The analysis in the HR diagram suggests a smaller than 
solar calibrated characteristic length scale for the MLT.

\subsection{Full spectrum of turbulence phenomenology}\label{42}

\begin{figure}[Ht]
\centering
\includegraphics[angle=90,width=8cm]{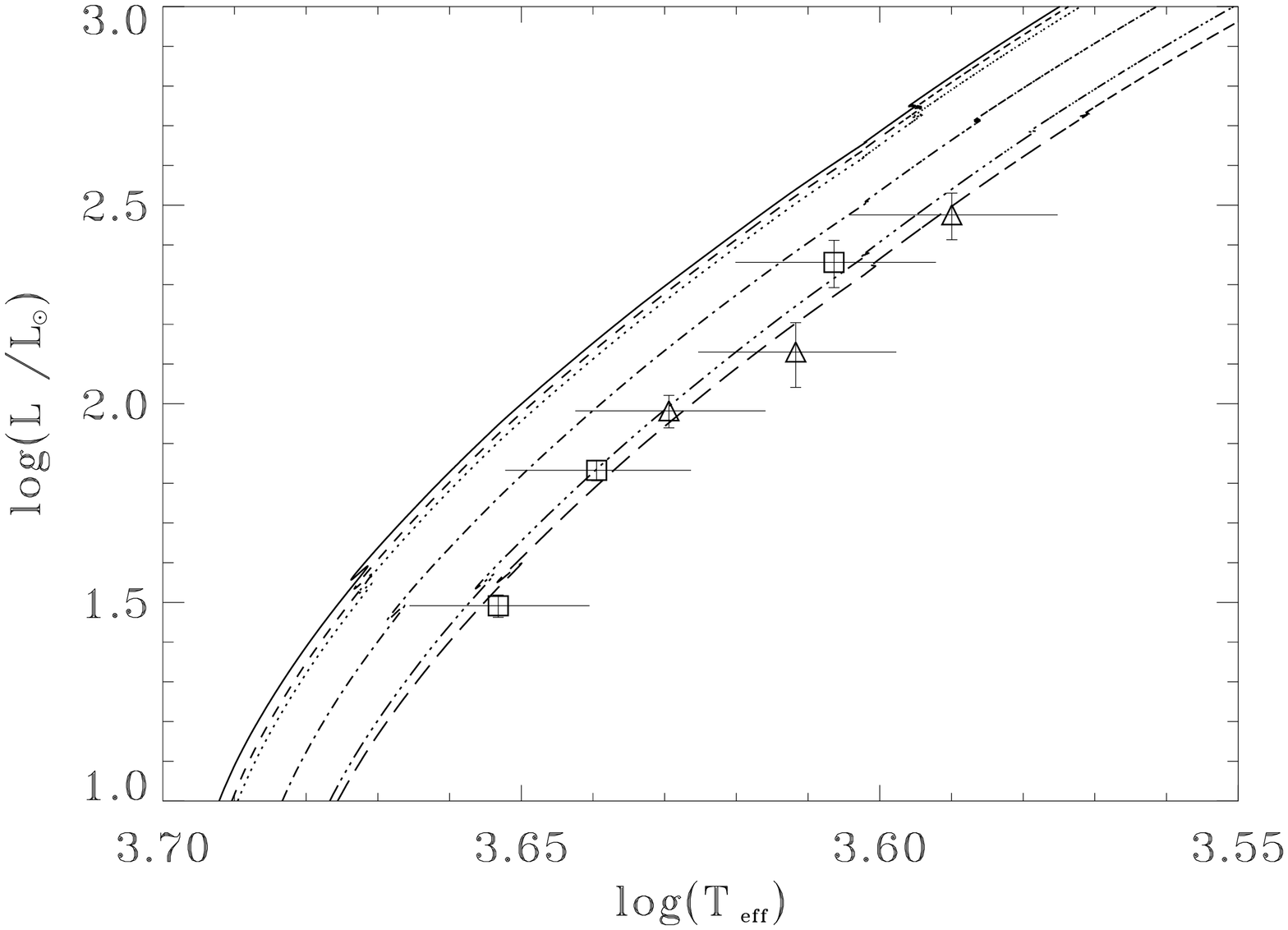}
\includegraphics[angle=90,width=8cm]{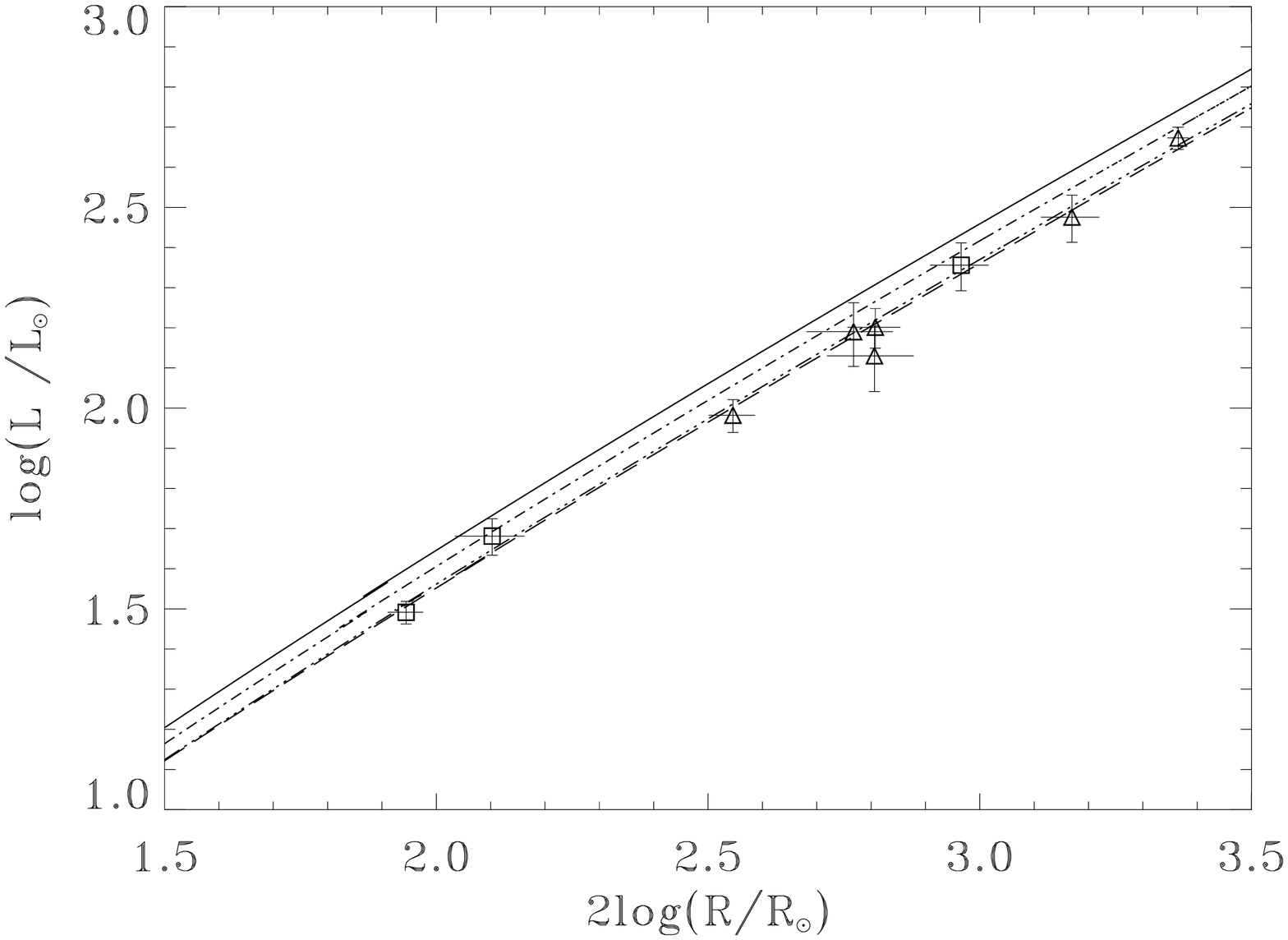}
\caption{{\bf Left:} Position of the six lower envelope stars of the sample with 
$\rm T_{eff}$ errorbars and various evolutionary tracks computed using
the MLT. See text for a description of the corresponding models.
{\bf Right:} Position of the nine stars of the sample with largest radii 
and errorbars. The various evolutionary tracks in overplot were computed using
the CGM. See text for a description of the corresponding models.}
\label{fig2}
\end{figure}

We will now perform a similar analysis as above but regarding the CGM 
phenomenology. Furthermore, instead of using $\rm T_{eff}$ we will use the 
interferometric radii. We therefore do not set ourselves in the HR diagram 
but in a luminosity vs. square of radius diagram 
(see Figure~\ref{fig2} right panel). Let's define 
$\rm \chi_2^2=\sum_{i=1}^{N}\frac{1}{N}[\frac{R_{mod}^{2}-R_{obs}^{2}}{\Delta R_{obs}^{2}}]^2$. 
Once again N is the number of objects considered. $\rm R_{obs}$, $\rm R_{mod}$ 
and $\rm \Delta R_{obs}$ are respectively the observed radius of an object, 
the radius of the model having the same luminosity as the object and the 
uncertainty on the observed radius.
Figure~\ref{fig2} right panel shows the nine stars with the largest radii at given 
luminosities. They were selected by considering the nine largest deviations 
to a linear fit of the whole sample in the luminosity vs. square radius 
diagram. 
The solid line track in Figure~\ref{fig2} right panel corresponds to a 0.95$\rm M_{\odot}$ 
star reaching $\rm 10^3 L_{\odot}$ at $\rm \approx$11.6 Gyr. This model has 
$\rm [Fe/H]=-0.17$, an helium fraction Y=0.2582 and the solar-calibrated value
$\rm \alpha_{cgm}=0.77$. This model corresponds to too small radii to fit the 
lower envelope of the RGB: $\rm \chi_2^2=7.2$. As in the case of the MLT, 
changes in mass and helium fraction are unable to significantly improve the 
agreement to the data and we will not discuss them. On the opposite if we 
decrease $\rm \alpha_{cgm}$ down to 0.62 we recover a good agreement to the 
observations. The three dotted-dashed line is the track of the model with 
$\rm \alpha_{cgm}=0.62$, 0.95$\rm M_{\odot}$, and $\rm [Fe/H]= -0.17$. This model 
reaches $\rm 10^3 L_{\odot}$ at 11.8 Gyr and has $\chi_2^2$=0.70. The long-dashed
line model has $\rm \alpha_{mlt}=0.62$, 1.17$\rm M_{\odot}$, and $\rm [Fe/H]=0$. 
It reaches $\rm 10^3 L_{\odot}$ at 6.9 Gyr and has $\chi_2^2$=0.40. The analysis 
in the luminosity radius diagram suggests a smaller than solar calibrated 
characteristic length scale for the CGM. 

\section{Conclusion}\label{sec5}

We modelled the Sun and local RGB stars in order to check if the change in 
surface conditions implies a change of the characteristic convection length 
scale $\rm \Lambda$ for two local treatments of convection: the MLT and the CGM.
In both cases we assume $\rm \Lambda=\alpha H_p$. Therefore we do not consider 
the original version of the CGM where the characteristic convection length 
scale is the distance to the boundary with the region stable with respect to 
convection.
At a given absolute luminosity $\rm \alpha$ tunes the position of the RGB in 
effective temperature and in radius. We have accurate data on absolute 
luminosities, effective temperatures and radii of all RGB stars we consider. 
The location of the coolest stars or equivalently the largest radii stars of 
the sample suggest a decrease in surface characteristic length scale with 
respect to its solar calibrated value. We have shown the decrease to be 
required for the MLT in the HR diagram and for the CGM in the luminosity 
radius diagram. However we could have inverted the diagrams with respect
to the convection 
treatments, the result would have been similar. The reader will find 
many more details in \cite{Piau11} where we also specifically address the three 
RGB stars of the sample with asteroseismic mass estimates. The combination 
of interferometric and asteroseimic data clearly opens up new perspectives 
in the understanding of stellar fundamental parameters and how they can be 
used to constrain stellar physics (\cite{Huber11}).

\begin{acknowledgements}
This work has been supported by LATMOS of Centre National de la
Recherche Scientifique and the Centre National d'Etudes Spatiales for the
scientific return of the Picard mission.
\end{acknowledgements}



%
\end{document}